\documentclass{aa}
\usepackage{graphics}
\usepackage{times}
\def\cm{\,{\rm cm}}
\def\cm3{\,{\rm cm^{-3}}}

\def\ion#1#2{\ifmmode \mbox{{\rm #1}}\,\mbox{{\sc #2}} 
        \else {\rm #1}$\,${\sc #2}
        \fi}
\def\HI{\ion{H}{i}}

\def\kms{\,{\rm {km\,s^{-1}}}}

\def\hmsd#1h#2m#3.#4s{                  
                      \relax
                      \ifmmode #1^{\rm h}\,#2^{\rm m}\,#3^{\rm s}
                               \hskip-0.39em.\hskip0.08em#4
                      \else \hbox{$#1^{\rm h}\,#2^{\rm m}\,#3^{\rm s}
                            \hskip-0.39em.\hskip0.08em#4$}
                      \fi
                     }
\def\dms#1d#2m#3s{                      
                  \relax
                  \ifmmode #1^\circ\,#2^{\prime}\,#3^{\prime\prime}
                  \else \hbox{$#1^\circ\,#2^{\prime}\,#3^{\prime\prime}$}
                  \fi
                 }
\def\degd#1.#2{                         
               \ifmmode {#1^{\hskip 0.05em\circ}\hskip-0.42em.\hskip0.08em#2}
               \else {#1$^{\hskip 0.05em\circ}\hskip-0.42em.\hskip0.08em$#2}
               \fi
              }
\def\mind#1.#2{                         
               \ifmmode {#1^{\hskip 0.05em\prime}\hskip-0.35em.\hskip0.05em#2}
               \else {#1$^{\hskip 0.05em\prime}\hskip-0.35em.\hskip0.05em$#2}
               \fi
              }
\def\secd#1.#2{                         
               \ifmmode {#1^{\prime\prime}\hskip-0.46em.\hskip0.12em#2}
               \else {#1$^{\prime\prime}\hskip-0.46em.\hskip0.12em$#2}
               \fi
              }
\def\mg{\relax                          
        \ifmmode ^{\rm m}
        \else $^{\rm m}$
        \fi
       }
\def\mgd#1.#2{                          
              \relax
              \ifmmode #1^{\rm m}
                       \hskip-0.55em.\hskip0.22em#2
              \else \hbox{#1$^{\rm m}
                    \hskip-0.55em.\hskip0.22em$#2}
              \fi
             }
\def\unitspace{\,}                      

\def\un#1{\ifmmode \unitspace{\rm #1} 
          \else $\unitspace$#1
          \fi}
\def\pun#1#2{\ifmmode \unitspace\mbox{\rm #1}^{#2} 
             \else $\unitspace$#1$^{#2}$
             \fi}
\def\mum{\ifmmode \unitspace\mu\mbox{\rm m} 
         \else $\unitspace\mu$m
         \fi}

\def\aua{{A\&A} }
\def\auas{{A\&AS} }
\def\apj{{ApJ} }
\def\aj{{AJ} }
\def\apjs{{ApJS} }

\def\araa{{ARA\&A} }
\def\pasp{{PASP} }
\def\pasj{{PASJ} }
\def\mnras{{MNRAS} }

\newcommand{\Figref}[1]{Figure~\protect\ref{#1}}

\begin{document}
 
\thesaurus{03         
          (11.01.2;
           11.09.1 NGC\,3079; 
           11.09.4; 
           11.14.1;
           13.09.1)
          }

\title{The obscured circumnuclear region of the outflow galaxy NGC\,3079}
 
\author{F.P.\ Israel\inst{1}
 \and   P.P.\ van der Werf\inst{1}
 \and	T.G.\ Hawarden\inst{2}
 \and   C.\ Aspin\inst{2, 3}
       }
 
\offprints{F.P.\ Israel}
\mail{israel@strw.leidenuniv.nl} 

\institute{$^1$ Sterrewacht Leiden, P.O.\ Box 9513, NL - 2300~RA Leiden,
             The Netherlands\\
$^2$ Joint Astronomy Centre, 660 N.~A'ohoku Pl., Hilo, Hawaii, 96720, USA\\
$^3$ Nordic Optical Telescope, Apartado 474, E - 38700, Santa Cruz de la
	     Palma, Canary Islands, Spain
}
 
\date{Received ... / Accepted ...}
 
\maketitle

\begin{abstract}
Images of the central region of the almost edge-on Sc galaxy NGC\,3079 in 
the $J$, $H$ and $K$-bands and in the $v=1{\to}0$ S(1) line of molecular 
hydrogen are presented. The inner few kiloparsecs of NGC\,3079 exhibit a 
large range of near-infrared colours caused by varying contributions from 
direct and scattered stellar light, emission from hot dust and extinction 
gradients. Our results show that interpretation of the observed light 
distribution requires high-resolution imaging in order to separate the 
different effects of these contributions.

The central 1$''$ (87\,pc) of NGC\,3079 suffers a peak extinction 
$A_{V}\sim\mg 6$. Its extremely red near-infrared colours require the 
additional presence of hot dust, radiating at temperatures 
close to 1000\,K. The least reddened eastern parts of the bulge require 
either a contribution of 20\% of light in the $J$-band from a younger 
population in a stellar bar or a contribution of 20--30\% from scattered 
starlight; scattered light from a nuclear source would require a less 
likely emission spectrum $S_{\nu}\propto\nu$ for that source.

The nucleus is surrounded by a disk of dense molecular material, extending 
out to a radius of about 300\,pc and with a central cavity. Bright H$_2$ 
emission and emission from hot dust mark the hole in the CO distribution 
and trace the inner edge of the dense molecular disk at a radius 
of 120\,pc. Less dense molecular gas and cooler dust extend out to radii 
of about 2\,kpc. 
In the inner few hundred parsecs of NGC\,3079, $\HI$ spin temperatures 
appear to be well below 275\,K and the CO-to-H$_{2}$ conversion factor has 
at most 5$\%$ of the Galactic value. An underabundance of H$_{2}$ with 
respect to CO is consistent with theoretical predictions for environments 
subjected to dissociative shocks, where reformation of H$_{2}$ is impeded 
by high dust grain temperatures. The overall molecular gas content of 
NGC\,3079 is normal for a late-type galaxy.

\keywords{galaxies: active -- individual: NGC\,3079 -- ISM --  nuclei;
infrared: galaxies}

\end{abstract}
 
\section{Introduction}

\begin{sloppypar}
NGC\,3079 is a bright and highly inclined late-type spiral galaxy (SB(s)c: 
De Vaucouleurs et al.\ 1976; Sc(s)II-III: Sandage $\&$ Tammann 1987) 
accompanied by the lesser galaxies NGC\,3073 and MCG\,$9{-}17{-}9$. The 
group distance is estimated to be 16\,Mpc for $H_{\rm 0}=75\kms\,{\rm 
Mpc}^{-1}$ (Irwin $\&$ Seaquist 1991). For consistency with Hawarden et 
al.\ (1995; hereafter HIGW), we will assume a distance of 18\,Mpc 
(Aaronson $\&$ Mould 1983) in the remainder of this paper. 
Some basic properties of NGC\,3079 are given in Table 1.
\end{sloppypar}

On both sides of NGC\,3079, strong lobes of radio continuum emission 
extend several kiloparsecs from the plane along the minor axis (De~Bruyn 
1977; Seaquist et al.\ 1978; Duric et al.\ 1983; Duric $\&$ Seaquist 1988), 
in the inner parts associated with filamentary H$\alpha$ and [$\ion{N}{ii}$] 
emission interpreted as the signature of a powerful outflow from the nucleus 
with velocities of up to $2000\kms$ in a cone of large opening angle
(Heckman et al.\ 1990; Filippenko $\&$ Sargent 1992; Veilleux et al.\ 1994).
In contrast to the radio lobes, the optical filaments are only seen at the 
eastern side of the disk, indicating that the western side suffers 
higher extinction. 

In the disk of NGC\,3079, radio continuum emission extended over 
$28''\times17''$ ($2.4 \times 1.5\,$kpc) surrounds a very 
compact (size about 1\,pc) radio core (Seaquist et al.\ 1978; Irwin $\&$ 
Seaquist 1988; Duric $\&$ Seaquist 1988; Baan $\&$ Irwin 1995) which 
agrees in position with an X-ray point source (Fabbiano et al. 1992) 
and strong H$_{2}$O masers (Henkel et al.\ 1984 
and references therein). Based on VLBI observations, Irwin $\&$ Seaquist 
(1988) argued that the outflow originates from a central compact object 
rather than from a more extended starburst region. 

\begin{table}
\caption[]{Properties of NGC\,3079}\label{tab.prop}
\begin{flushleft}
\begin{tabular}{ll}
\hline
Type$^{a}$     	 			& Sc(s)II--III \\
R.A. (B1950)$^{b,c}$ 	 		& $\hmsd 09h58m35.0s$  \\
Decl.(B1950)$^{b,c}$        		& $\dms 55d55m15s$ \\
$v_{\rm hel}$$^{b,d}$  			& 1125 ($\HI$) -- 1145 (CO) $\kms$ \\
Distance $D^{e}$          		& 18\,Mpc \\
Inclination $i^{b}$ 			& $\degd84.5$ \\
Position angle $P^{b}$     		& $\degd166.5$ \\
Luminosity $L_{\rm B}^{f}$  		& $3.4 \times 10^{10}\,L_{\odot}$ \\
Luminosity Nucleus $L_{\rm FIR}^{g}$ 	& $0.7 \times 10^{10}\,L_{\odot}$ \\
Scale           			& 11.5$''$/kpc or 87\,pc/$''$ \\
\hline
\end{tabular}
\end{flushleft}
$^{a}$ RSA (Sandage $\&$ Tammann 1987) \\
$^{b}$ $\HI$ observations, Irwin $\&$ Seaquist (1991) \\
$^{c}$ Baan $\&$ Irwin (1995) \\
$^{d}$ Sofue $\&$ Irwin (1992); Braine et al. (1997) \\
$^{e}$ Aaronson $\&$ Mould (1983), for $H_{0} = 62\kms$\,Mpc$^{-1}$ \\
$^{f}$ Fabbiano et al.\ (1992) \\
$^{g}$ Hawarden et al.\ (1995) \\
\end{table}

In NGC\,3079, CO emission is concentrated in the centre (Irwin $\&$ Sofue 
1992; Sofue $\&$ Irwin 1992). Out to radii of $\secd 5.4$ (470\,pc), the 
molecular gas exhibits solid body rotation with a maximum rotational 
line-of-sight velocity of $330\kms$. The steep velocity gradient and the 
finite angular resolution of about 4$''$ create the false impression 
of an unresolved CO emission peak at the nucleus. However, 
position-velocity maps (in particular Fig.~5b in Sofue $\&$ Irwin 1992) 
show the existence of a small nuclear hole in the CO distribution, similar 
to the situation found in the equally highly inclined galaxy NGC\,253 
(Israel et al.\ 1995). 

Duric $\&$ Seaquist (1988) explained the then-observed phenomena
with a model in which the observed radio structures result from a strong 
nuclear wind focussed into a bipolar outflow by a dense circumnuclear disk. 
This model was supported by HIGW on the grounds that the observed 
properties of the centre of NGC\,3079 cannot be explained by a 
(circum)nuclear starburst, but rather point to the existence of an 
active nucleus vigorously interacting with its gaseous surroundings. 
HIGW conclude that the H$_2$ vibrational line emission is not excited 
by X-rays or UV photons. Instead, they argue that kinetic energy 
of fast shocks generated by wind impact on the molecular gas disk is 
converted into H$_{2}$ line mission, with the low efficiency expected for 
such a mechanism, and that the extended mid-infrared emission from NGC\,3079 
arises from shock-heated dust. 

HIGW did not obtain images of the H$_{2}$ distribution, and only barely
resolved its emission. Because NGC\,3079 is seen almost edge-on, 
its centre suffers considerable extinction (Forbes et al.\ 1992; 
Veilleux et al.\ 1994), leading to some uncertainty in the near-infrared 
luminosities discussed by HIGW. In order to verify the conclusions reached 
by them, a further investigation of the nuclear H$_{2}$ emission and the 
properties of the central region of NGC\,3079, by high resolution imaging of 
the $J$, $H$ and $K$-band continuum emission
as well as the $v=1{\to}0$ S(1) H$_{2}$ line emission, was deemed desirable.

\section{Observations}

\subsection{Broad-band images}\label{sec.JHK}

The $J$, $H$ and $K$-band images were obtained on 1994 May 18at UKIRT under 
non-photometric conditions using IRCAM3, the UKIRT near-infrared imaging 
camera, through standard $J$, $H$ and $K$-band filters (1.25, 1.65 and 2.2 
$\mu$m respectively), on a Santa Barbara Research Corporation InSb array of 
256$\times$256 pixels, with a pixel scale of $\secd0.286$ on the sky, 
resulting in a $73''$ field. 
The image of a foreground star at $\Delta \alpha = +15''$, $\Delta
\delta = +2''$ was used to align the different frames. 
The flux calibration of the 
individual images was derived from the synthetic aperture photometry
by Forbes et al.\ (1992), using their largest ($12''$) aperture values 
in order to minimize the effects of positional differences. Consistency 
checks using their smaller aperture values revealed deviations of up to 
3\%, which can be regarded as the uncertainty of our flux calibration 
procedure. The effective resolution of the final images, measured from 
the foreground star referred to above, is $\secd1.0$

\subsection{Molecular hydrogen vibrational emission line images}

\begin{sloppypar}
The H$_2$ $v=1{\to}0$ S(1) images were obtained in May 1992 using the 
near-infrared Fabry-Perot imaging spectrometer FAST (Krabbe et al.\ 1993) 
at the Cassegrain focus of the 4.2\,m William Herschel Telescope (WHT) 
at Roque de los Muchachos at La Palma, Spain. The FAST camera used a 
Santa Barbara Reserarch Corporation 58$\times$62 InSb array with a 
pixel scale of $\secd 0.5$ and a field of about $30''\times30''$.
Dispersion was provided by a Queensgate scanning Fabry-Perot interferometer 
with a spectral resolution ($\lambda{/}\Delta\lambda$) of 950 at $2.13\,\mu$m 
(corresponding to a velocity resolution of 315 km s$^{-1}$), used in tandem 
with a $\lambda{/}\Delta\lambda$ = 45 cold circular variable filter (CVF)
as order sorter. The seeing was about $\secd 1.2$. Because the S(1) line 
has a full width at 20$\%$ intensity of $630\kms$, we took line images not 
only at the systemic velocity but also at velocity offsets of $\pm282\kms$,
as well as line-free continuum images at velocity offsets $\pm707\kms$, i.e.
at five velocity settings in total. Several sets of images were obtained 
with exposure times of $150\,$sec each. Sky frames were obtained at a 
position 6$\arcmin$ east of the nucleus. 
After subtraction of the dark current, the individual frames were 
flatfielded and sky-subtracted. The resulting line-plus-continuum 
images were corrected for atmospheric transmission and instrumental 
response with the use of the standard stars HR\,3888 and HR\,4550. 
Finally, the mean of the continuum on either side of the line (velocity 
offsets $\pm707\kms$) was subtracted. The resulting line images 
were co-added, yielding line flux maps with a total integration time of 
1650\,sec at the systemic velocity and 750\,sec each for the images offset 
in velocity by $\pm282\kms$.

\begin{figure*}[tph]
\resizebox{\hsize}{!}{\includegraphics{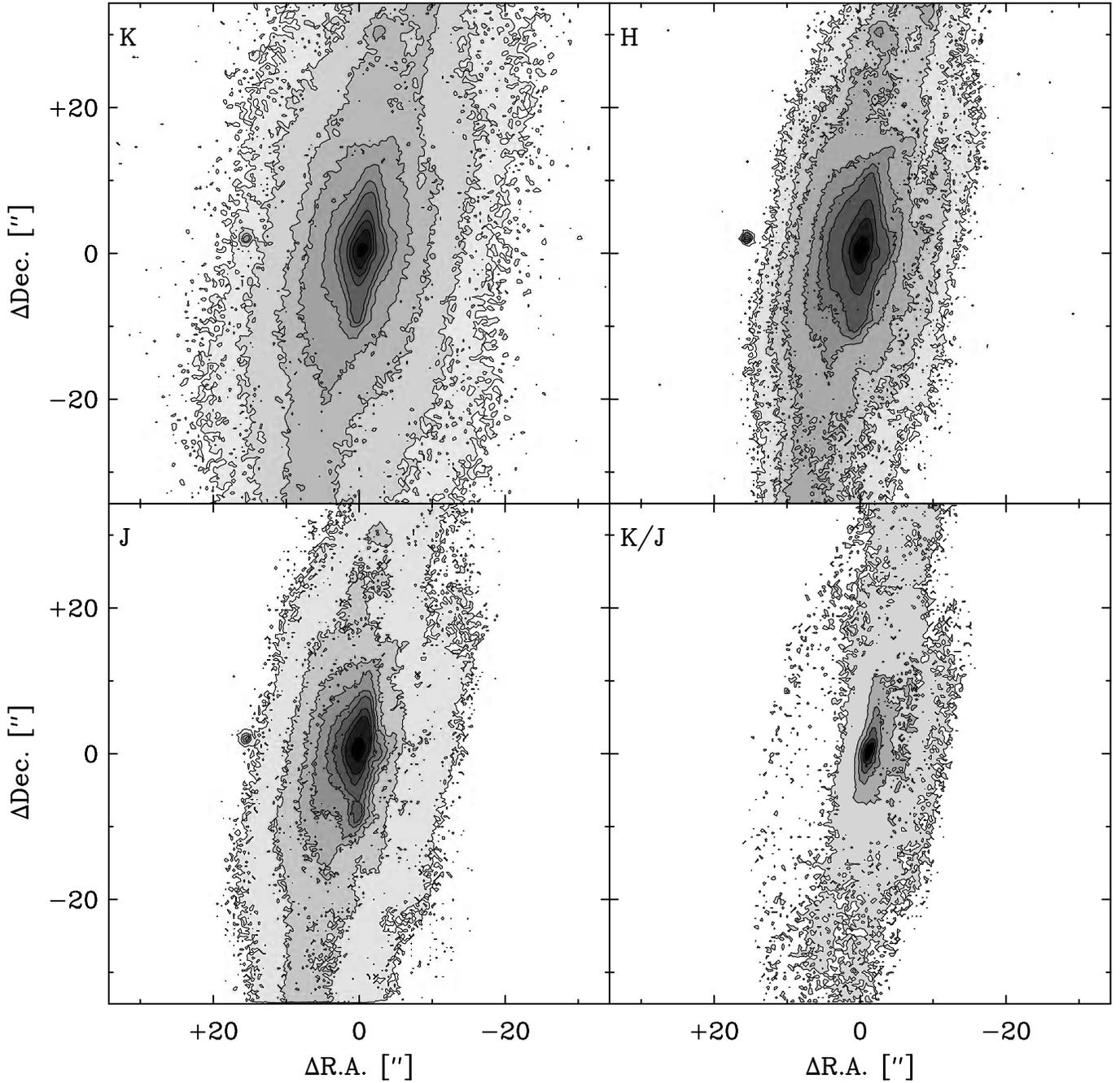}}
\caption[]{
Near-infrared continuum images of the central $\mind 1.14$ of NGC\,3079 
in the $K$-band (top left), the $H$-band (top right) and the $J$-band 
(bottom left).
Contour levels are 0.1, 0.2, 0.4, 0.6, 1, 1.5, 2, 4, 6, 10 and 20
 $\times10^{-5}$\,W\,m$^{-2}$\,$\mu$m$^{-1}$\,sr$^{-1}$ ($K$-band), 
0.5, 0.75, 1, 1.5, 2, 2.5, 5, 10 and 20 
$\times10^{-5}$\,W\,m$^{-2}$\,$\mu$m$^{-1}$\,sr$^{-1}$  ($H$-band), 
0.75, 1.5, 2.25, 3, 4, 5, 7.5, 10 and 20 
$\times10^{-5}$\,W\,m$^{-2}$\,$\mu$m$^{-1}$\,sr$^{-1}$  ($J$-band).
The frame at bottom right shows the $K{/}J$-band flux ratio, with 
contours at ratios of 0.3 to 1.5 in steps of 0.2; the reddest regions 
are darkest. Positions are relative to the $K$-band peak.
}
\label{fig.KHJ}
\end{figure*}

We also obtained images at the wavelengths of Br$\gamma$ ($2.1655\,\mu$m)
and [$\ion{Fe}{ii}$] ($1.6435\,\mu$m), centered on the systemic velocity 
under good conditions (seeing $\secd0.8$). Total integration times
were 2000\,sec and 1500\,sec respectively. Resulting r.m.s.\ noise figures 
are $1.5\times10^{-8}\,$W\,m$^{-2}$\,sr$^{-1}$ for Br$\gamma$ and 
$1.5\times10^{-7}\,$W\,m$^{-2}$\,sr$^{-1}$ for [$\ion{Fe}{ii}$].

No [$\ion{Fe}{ii}$] emission was detected, but the Br$\gamma$ 
image shows weak emission centred on the nucleus, just 
above the noise level and extended over about 3$''$. The total 
measured flux of about $5\times10^{-18}\,$W\,m$^{-2}$, is reasonably
consistent with the marginal detections by HIGW: 
$F({\rm Br}\gamma)=2.9\pm1.2\times10^{-17}\,$W\,m$^{-2}$ 
in a large aperture CVF spectrum and
$F({\rm Br}\gamma)=1.2\pm1.0\times10^{-17}\,$W\,m$^{-2}$ 
in a 3$''$ UKIRT CGS4 aperture. Taken at face value, these 
rather uncertain data suggest the presence of weak and diffuse ionized 
hydrogen extended over the inner $1-2\,$kpc, and somewhat concentrated 
near the nucleus.
\end{sloppypar}

\section{Results}

\subsection{Broad-band images}

The $J$, $H$, and $K$-band images and a $J/K$ image are reproduced 
in Fig.\, 1.

The differences between the broad-band images primarily reflect the
wavelength-dependent effects of extinction. The strong east-west 
asymmetry, most clearly seen in the $J$-band image, is caused by dust 
extinction in the western half of the galaxy. With increasing wavelength, 
the light distribution becomes more symmetrical and more pronouncedly 
peaked, although symmetry is not perfect even at $K$-band, suggesting 
obscuration of the nucleus even at this wavelength. The $J$ and $K{/}J$ 
images indicate the presence of at least two dust lanes: one cuts off 
the bulge light, the second betrays its presence by bays of reduced 
emission in the $J$-band image, corresponding to an extended reddened 
zone in the $K{/}J$ image about 10$''$ west of the nucleus, parallel to 
the midplane. The relatively well-defined western edge in the $J$-band 
image may represent a third dust lane, seen as a thin layer of enhanced 
reddening between the other two dust lanes in the $K{/}J$ image. 
The north-south asymmetry apparent in the images suggests higher 
extinction north of the nucleus than south of it. East of the nucleus, 
the bulge light appears to suffer less extinction, and a bulge with a 
peanut-shaped light distribution is apparent (Shaw et al.\ 1993). 
The asymmetries along the major and along the minor axes are further 
illustrated by the near-infrared colour profiles shown in Fig.\,2.

\begin{figure}[t]
\resizebox{\hsize}{!}{\includegraphics{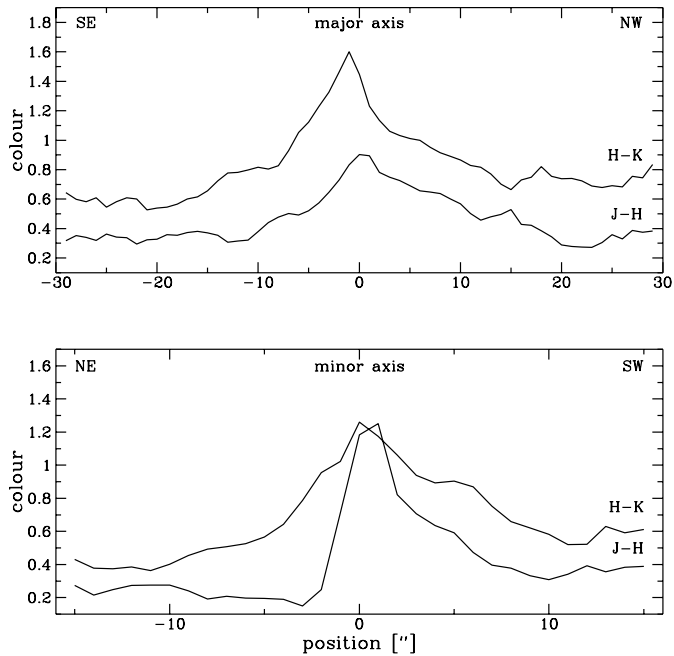}}
\caption[]{Near-infrared colour profiles 
in $1''$ apertures along the major axis
(top panel) and along
the minor axis (bottom panel) of NGC\,3079. Positions are with respect to
the peak of the $K$-band peak.  
}\label{fig.profiles}
\end{figure}

Lawrence et al.\ (1985) noted a displacement of the 10$\,\mu$m and radio 
peak (presumably marking the true nucleus) from the {\it visual} peak by 
5$''$ in a northwestern direction. The near-infrared images at 
wavelengths intermediate between 10$\mu$m and the visual show the same 
effect but, as expected, to a lesser degree. Going from $J$ to $K$, the 
intensity peak shifts to the west by $\secd0.9$. 

All three images (especially the $J$-band image) show a secondary maximum 
8$''$ south of the main peak. The lower resolution 0.8\,mm continuum 
map by HIGW shows the central source to be elongated in this direction, 
and the radio continuum maps by Duric et al. (1983) and Baan $\&$ Irwin 
(1995) show enhanced emission at this position, with a $1.4-5.0$\,GHz 
radio flux density of about 0.5\,mJy. The apparently flat radio 
continuum spectrum, implying mostly thermal emission, is consistent 
with the intensity of the bright H$\alpha$+[$\ion{N}{ii}$] knot visible 
in Fig.\,1 of Veilleux et al.\ (1994) at this position. The object thus 
appears to be a massive star forming region; its location, especially 
in the H$\alpha$+[$\ion{N}{ii}$] image, suggests that it is part of a 
spiral arm, and at a projected distance from the nucleus of about 
700\,pc. Its radio intensity implies the presence of a few hundred 
OB stars. The $J$-band image contains a few more, weaker peaks farther 
out in the disk of NGC\,3079 (e.g., at $-2.5$, $+30$), which may 
likewise represent concentrations of luminous stars.

\begin{figure}[t]
\vspace{1.0cm}
\caption[]{H$_{2}$ $v=1{\to}0$ S(1) emission line surface brightness 
images of the centre of NGC\,3079.
``Channel'' maps represent emission integrated over $315\kms$ velocity 
intervals, at central
velocities (relative to the systemic velocity of NGC\,3079) indicated in the 
figures. The sum of the three channel maps is at bottom
right. Contour values are 2, 4, 7, 10, 13, 16 and 19 in units of 
10$^{-8}$\,W\,m$^{-2}$\,sr$^{-1}$ for the channel maps, and 
4, 8, 12, 18, 24, 30, 36, 42 and 48 again in units of 
10$^{-8}$\,W\,m$^{-2}$\,sr$^{-1}$ 
for the map of integrated emission. Positions are relative 
to the peak position of the integrated
H$_2$ emission. The cross marks the position 
of peak broad-band 2.1\,$\mu$m emission. }
\label{fig.H2channels}
\end{figure}

\subsection{H$_{2}$ images}

In Fig.\,3 we show the line surface brightness maps of 
the H$_{2}$ emission from the centre of NGC\,3079. As the velocity 
separation is only 10$\%$ less than the velocity resolution, the 
``channel'' images are largely independent. The H$_{2}$ distribution 
has a relatively sharp western edge and a more wispy eastern boundary.
The peak of the total H$_{2}$ emission is located $\secd0.9$ north 
of the $2.13\mum$ continuum peak marked by a cross in Fig.\,3. 
This offset is real and accurate, since both 
the continuum and line images are extracted from the same data set. 
In Fig.\,4 we further illustrate this offset by plotting the outline 
of the total H$_{2}$ emission on a contour map of the 2.13\,$\mu$m 
continuum distribution.

The peak of the H$_2$ emission is found at accurately the same position in
all of the three channels maps of Fig.\,3. Because the emission line is 
widest at this position, we identify it with the dynamical centre of the 
H$_2$ emission. This position thus likely marks 
the obscured nucleus of NGC\,3079. The position offset of the $K$-band 
continuum nucleus towards the south indicates that extinction still 
plays a role in this wavelength region and that the H$_2$ emission may
suffer less extinction than the continuum.

\begin{sloppypar} 
The H$_{2}$ distribution in Fig.\,3 can be schematically 
described by a bright central component superposed on more extended 
emission. The bright central component measures 3$''\times$2$''$ 
(260$\times$175\,pc) in the integrated H$_{2}$ map. In the central channel, 
this component is elongated with diameters of $3'' \times\secd1.4$ 
at position angle ${\rm PA} = 147\pm 5^{\circ}$, nominally deviating
from the extended disk position angle. 
At the outlying velocities, the peak surface brightness of this component 
has dropped by a third and the shape is more circular with a diameter 
of $\secd1.9$. Thus at the outlying velocities, the central bright 
H$_{2}$ component is {\it less} extended in the plane of the galaxy 
and {\it more} extended perpendicular to the plane than at the systemic 
velocity. This result will be discussed further in Sect.\,4.2.2.
\end{sloppypar}

\begin{figure}[t]
\vspace{1.0cm}
\caption[]{The outline of the compact total H$_{2}$ $v=1{\to}0$ S(1) line 
emission plotted on top of a contour map of the more extended 2.1\,$\mu$m 
continuum emission. H$_{2}$ contour values are 5, 20 and 40 in units of 
$10^{-8}\,$W\,m$^{-2}$\,sr$^{-1}$; continuum contours are 10, 20, 30, 50, 
75, 100, 150, 200, 300, 400, 600, 800 and 1000 in arbitrary units.
}
\label{fig.H2total}
\end{figure}

The more extended emission is fainter. In integrated H$_{2}$ it measures 
$6''-7''$ along ${\rm PA} = 157^{\circ}$, closer to the 
position angle ${\rm PA} = \degd 166.5$ of the disk (Table\,1)
In the perpendicular direction, its extent is hard to determine because 
of the dominating presence of the central component. The channel maps 
show the extended component to be rotating, with the south receding and 
the north approaching.

The integrated intensity of the H$_{2}$ $v=1{\to}0$ S(1) line emission 
in Fig.\,3 is $7\times10^{-17}\,$W\,m$^{-2}$, 
which is about two thirds of the line strength found by HIGW from 
spectrophotometry.

\section{Analysis and discussion}

\subsection{Continuum colours}

As noted by HIGW, the stellar absorption spectrum of the nucleus at 
$2.3\,\mu$m and the overall spectral shape of the near-IR continuum  
suggest that out to $5\,\mu$m the radiation from the centre of the
galaxy is dominated by emission from late-type (super)giants in the bulge. 
Near-infrared photometry was published by Lawrence et al.\ (1985), 
Willner et al.\ (1985) and Forbes et al.\ (1992). The colours derived
by Forbes et al.\ (1992) suggest relatively low extinction, but this 
must be interpreted as a lower limit because these authors 
used fairly large apertures, and, more importantly, first aligned 
the peaks of their $J$, $H$ and $K$-band images, while we have noted 
in Sect.\,3.1 that these are actually displaced from one another. 
The colours measured by Willner et al.\ (1985) suggest that the central 
6$''$ of NGC\,3079 suffers a high extinction $A_V\sim 7-11\mg$ 
(their Figs.~2 and 3). HIGW analysed the available data and 
concluded to a ``best'' value $A_V = \mgd 7.2\pm\mgd0.9$ for the 
central 6$''$.

We have used our accurately aligned $J$, $H$ and $K$-band images 
to produce the two-colour diagram shown in Fig.\,5,
and we analyse the colours of the various regions in NGC\,3079 in the 
following sections. The curves marked ``screen'' and ``mixed'' in this figure 
indicate the effects upon the observed colours of extinction by respectively
foreground dust and dust uniformly mixed with the stellar population. In 
the former case, $I_{\rm obs}=I_0 \rm e^{-\tau}$, 
while in the latter case,
$I_{\rm obs}=I_0 (1-\rm e^{-\tau})/\tau$, where 
$I_{\rm obs}$ and $I_0$ are respectively 
the observed and intrinsic intensities. A Galactic extinction curve is assumed.

\subsubsection{The outer bulge}

The $H-K$ colour observed from the outer bulge (i.e., positions 
4$''$ or more east of the disk) agrees well with that of typical 
bulges, but the $J-H$ colour is about $\mgd 0.20$ bluer.
These results are similar to those found by HIGW in a $6''$ 
aperture. From the observed $K$-band absorption features, HIGW 
determined that the mean spectral type of the bulge population 
dominating the $K$-band light is M0III, reasonably consistent 
with the ``bulges'' zeropoint in Fig.\,5 and with the 
observed $H-K$ colour of the eastern bulge in NGC$\,$3079. 
This leaves the blueness of the eastern bulge in $J-H$ to be 
explained. HIGW note that blue colours also prevail at wavelengths
shorter than $1\,\mu$m and that these colours point to the 
contribution of young stars at $1\,\mu$m and shorter wavelengths. 
There are several possibilities.

\begin{figure*}[th]
\resizebox{12cm}{!}{\includegraphics{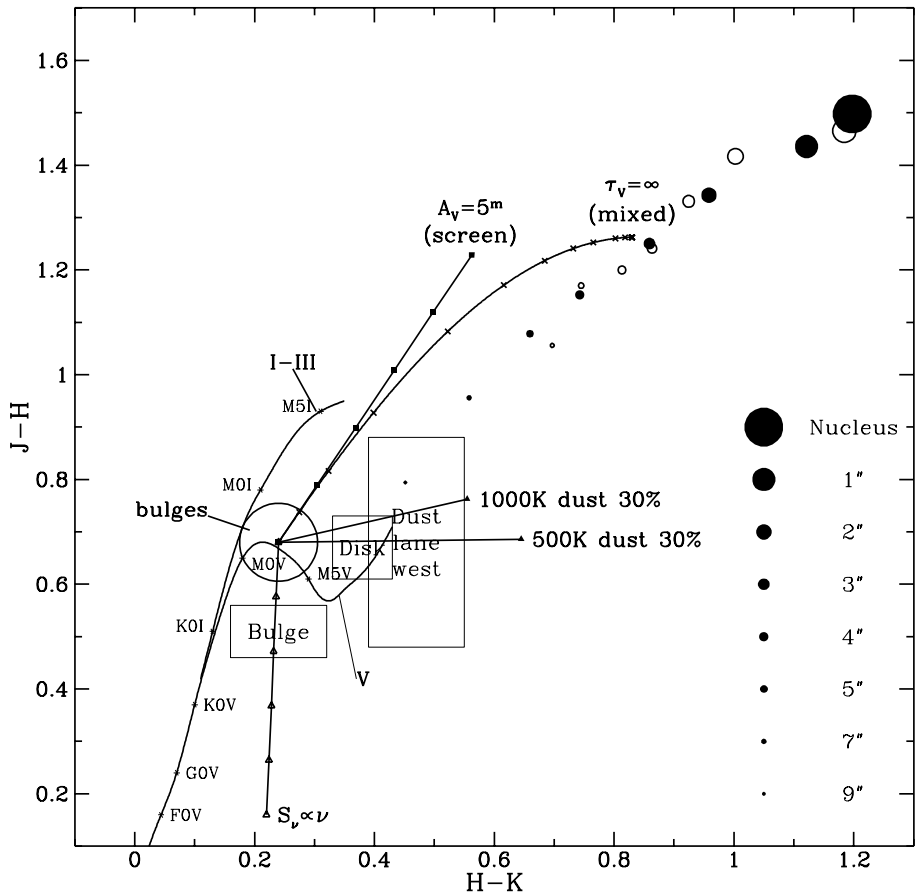}}
\hfill
\parbox[b]{55mm}{
\caption[]{Two-colour diagram of NGC\,3079 near-infrared emission. 
Circles of various sizes indicate colours in $1''$ apertures along 
the disk of NGC\,3079 (assumed position angle $\degd166.5$), displaced 
by the indicated amounts from the nucleus, assumed to be at the 
dynamical centre of the H$_2$ emission. Open circles represent the 
disk north of the nucleus, while filled circles denote positions 
south of the nucleus. In addition, rectangles indicate the colour 
ranges found in the eastern bulge well away ($>4''$) from the disk 
(labeled ``Bulge''), in the disk well north and south of the nucleus 
(``Disk'') and in dust lane west of the nucleus (``Dust lane west''). 
For comparison, the colours of unreddened bulges (Kuchinski \& Terndrup 
1996) are indicated by the large open circle (``bulges''). Solid 
curves identify the observed colours of Galactic main sequence stars 
of approximately solar abundance (curve marked ``V'') and of (super)giants 
(curve marked ``I--III'') of the indicated spectral types (Koornneef 
1983). Additional solid lines show the effects upon the 
observed colours caused by the presence of emission from hot (500 or 
1000\,K) dust contributing 30\% of the observed $K$-band flux, 
extinction by foreground dust for $A_V=1\mg$ to $5\mg$ in steps of 
$1\mg$ (``screen''), extinction by dust mixed with the stars with 
$V$-band opacities $\tau_V$ of 0, 1, 2.5, 5, 10, 15, 20, 25, 30, 40,
50 and $\infty$ (``mixed''), and the presence of a blue power law 
($S_{\nu}\propto\nu$) contributing 0, 20, 40, 60, 80 and 100\% of the 
observed $J$-band emission.
}
\label{fig.twocol}}
\end{figure*}

\begin{enumerate}
\item The blueness of the eastern bulge could be due to a contribution 
of scattered light from a nuclear source. As shown in Fig.\,5, 
scattered power law emission with $S_{\nu}\propto \nu^{\alpha}$ 
around $1\,\mu$m would require $\alpha\sim1$, and would have to 
contribute about 35\% of the $J$-band emission of the eastern bulge. 
However, the power law nuclear emission found in active galaxies 
typically has a much lower exponent around $1\mum$, e.g., $\alpha\sim-1$ 
in the sample of 34 Seyferts of Kotilainen et al.\ (1992), where no 
object has $\alpha>0$. 
\item The bulge of NGC$\,$3079 is peanut-shaped (Fig.\,1).
This is usually taken as the signature of a stellar bar (Combes \& 
Sanders 1981; Kuijken \& Merrifield 1995). Bars frequently harbour 
enhanced star formation, and the bar potential allows young stars 
in the bar to migrate into the peanut-like features. If this 
explanation is correct, these stars would have to contribute about 
20\% of the $J$-band light on the assumption of a characteristic 
spectral type A0V\@. 
\item Alternatively, the relative blueness of the the eastern bulge 
could be due to a contribution by scattered starlight. For instance, 
a reddening corresponding to $A_{V}$ = 2.7 mag (i.e. 50\% light loss 
at J) combined with a contribution of $30\%$ scattered light at J 
will match the observed colours to the typical bulge colour. If 
instead we assume extinction by dust mixed with stars, $\tau _{V}=5$
(75\% light loss at J) and a slightly smaller contribution by 
scattered light (20\% at J) will also explain the observed colours. 
Scattering of blue light from the young disk population, contributing 
about 20\% of the observed $J$-band light likewise is a viable 
alternative, and consistent with the overall optical blueness, 
$(B-V)_{0} = 0.46$ (RC2), of this edge-on galaxy. For various 
reasons (see Kuchinski $\&$ Terndrup 1996) these are only crude 
estimates; nevertheless they all suggest that scattered light 
contributing of the order of 20--30\% to the observed J-band
emission of the outer bulge is required to explain its near-infrared 
colours.
\end{enumerate}

\subsubsection{Disk and western dust lane}

In the disk away from the central part of NGC\,3079, the colours
of the stellar population must correspond to a mean spectral type 
earlier than those indicated by the ``bulges'' zeropoint, i.e.
close to those of the ``main-sequence line'' in Fig.\,5.  
If, for instance, a mean spectral type A0 is assumed for the intrinsic 
``disk'' colours, a ``screen'' reddening corresponding to $A_V=6\mg$ 
would be required; the dust lane would require a mean spectral type 
of early B and $A_V=8\mg$. Since recent star formation is associated 
with dust, this result is reasonable. Alternatively, the ``mixed'' 
extinction model may be a more adequate representation of the 
relative distributions of stars and dust in the disk of the galaxy. 
This model suggests a mean spectral type of late F or early G, but 
also rather high values $\tau_V=10-20$ for the extinction optical depth.
In reality, a combination of ``screen'' and ``mixed'' extinction is 
probably appropriate, and intermediate values for both spectral type 
and visual light loss are obtained by varying the relative importance 
of ``mixed'' and ``screen'' extinction. The present data do not allow 
us to draw a firm conclusion as to which combination is preferred.

\subsubsection{Stellar emission and $K$-band excess in the central kiloparsec}

In Fig.\,2a, we have shown the $J-H$ and $H-K$ colours in 1$''$ 
diameter apertures along the major axis of NGC\,3079. A number of these 
positions are also marked in Fig.\,5,. Towards the nucleus, 
the colours redden rapidly, reaching peak values $(J-H)=\mgd1.5$, 
$(H-K)=\mgd1.2$ at the nuclear arcsec$^{2}$ (87$\times$87\,pc). As 
Fig.\,5 shows, the ($H-K$) colour within $3''$ from the 
nucleus is far too red to be explained by either the ``screen'' or 
the ``mixed'' extinction curves in Fig.\,5, so that 
excess emission must be present in the $K$-band. The amount of excess 
$K$-band emission implied by the near-infrared colours depends, however, 
on the choice of extinction model. We argue here that the ``screen'' 
model is appropriate, for the following reasons.
\begin{enumerate}
\item 
First and foremost, the fact that NGC\,3079 is very nearly edge-on 
($i=\degd 84.5$) and displays prominent dust lanes implies that the 
nuclear region must undergo a large amount of foreground (``screen'') 
extinction.
\item
The occurrence of fast outflows from the nuclear region requires a 
significant central volume swept clear of gas and dust. Indeed, 
Sofue $\&$ Irwin (1992) have noted the presence of such a ``hole'' 
with a diameter of about $\secd 2.5$ in the CO distribution. Stars 
in this central cavity will therefore undergo only foreground 
extinction. 
\begin{sloppypar}
\item
We may use the 0.8\,mm continuum measurements by HIGW to estimate 
the column density of the emitting dust, and hence its visual optical 
depth and extinction. From Hildebrand (1983) and Savage $\&$ Mathis 
(1979) we derive for a dust emissivity proportional to $\lambda^{-1.5}$ 
the relation $A_V=2\times10^{4}\,\tau_{0.8}$, with a factor of about 
two uncertainty but independent of actual dust-to-gas ratios. In a 
16$''$ aperture, HIGW determined an 0.8\,mm flux density of 
$0.35\,$Jy for the unresolved central source, implying $\tau_{0.8} = 
6.4\times10^{-4}$ for a dust temperature $T_{\rm d} = 35\,$K (see
e.g. Braine et al. 1997). Since the emitting material surrounds the 
nucleus, only half of it will contribute to the extinction. Thus, 
the submm result implies an extinction $A_V = \mgd6.5\pm\mgd0.8$ or 
optical depth $\tau_V=6.0\pm0.7$. \Figref{fig.twocol} shows that 
this value of $\tau_V$ would not nearly produce the required reddening 
if the dust were mixed with the stars. Hence a dominant foreground 
extinction component is required.
\end{sloppypar}
\item
With an observed $K$ magnitude of $\mgd12.7$ in the central arcsecond 
and a distance modulus $(m-M) = \mgd 31.3$, the absolute $K$ magnitude 
becomes $\mgd-18.6$, not corrected for extinction. Large reddening 
corrections, as would result from the use of the ``mixed'' reddening 
curve, must thus be considered unlikely.
\item
If the blue ($J-H$) colour of the eastern bulge is due to a young stellar 
population, the reddening vectors in Fig.\,5 should begin 
in the rectangle marked ``bulge''. The reddened points along the disk 
then lie very closely to the screen extinction model, indicating a 
gradually increasing foreground extinction, as expected for a circular 
disk seen edge-on. In contrast, in this case the mixed extinction model 
has great difficulty explaining the gradual reddening towards the 
nucleus, producing colours that are too red in ($H-K$).
\item
Finally, Fig.\,5 shows that for dust mixed with the stars, 
the nuclear colours cannot be reproduced even in the limit of infinite 
$\tau_V$. Furthermore, the remaining colour difference $J-H$ = $\mgd0.23$, 
($H-K$) = $\mgd0.37$ cannot be reproduced by dust emission for any temperature 
below the dust sublimation temperature. This problem is even exacerbated 
if the bulge contains a young stellar population.
\end{enumerate}
All of these arguments indicate that the extinction towards the nuclear
region is dominated by foreground material. We will therefore in the 
following assume foreground extinction exclusively. 

\subsection{The central region of NGC\,3079}

\subsubsection{Hot dust}

The $JHK$ colours alone are insufficient to uniquely separate the 
effects of extinction and dust emission, as the observed points in
Fig.\,5 may be reached by different tracks. They do 
constrain, however, the range of admissible parameters.

On the one hand, Fig.\,5 shows that the screen extinction 
limit for the central position is $A_V<\mgd7.5$ at arbitrarily low 
$T_{\rm d}$. For the other positions, the limit is even lower. On the 
other hand, for a dust emissivity $\propto\lambda^{-2}$, dust-mixing 
curves with temperatures $T_{\rm d}>1000\,$K yield colours too blue
in ($H-K$) to explain the observed colours of the central position 
for any extinction, i.e. they pass to the left of the central point 
in Fig.\,5. As that figure shows, somewhat higher dust 
temperatures are in fact allowed for the other positions, but in the 
absence of starburst activity, we consider it unlikely that the 
projected central 250~pc is cooler than its surroundings. We thus find 
that $T_{\rm d}<1000\,$K for the center of NGC~3079. The 
photometry by Lawrence et al. (1985) in a 6$''$ aperture provides an 
additional constraint. Their observed ($K-L$) and ($K-M$) colours are very 
close to reddening lines originating in the ``bulge'' point. A 
non-negligible dust contribution, required by Fig.\,5, 
is therefore possible only for relatively high dust temperatures 
that likewise bring the dust-mixing line close to the reddening line. 
Taking into account the fact that Lawrence et al. (1985) were pointing 
off the true nucleus, their Fig. 1a suggests that, were they properly 
centered, the $L$ and $M$ magnitudes may be lower by at most $\mgd0.14$ and 
$\mgd0.43$ respectively. The resulting colours, ($K-L$) = $\mgd0.75$ and 
($K-M$) = $\mgd0.79\pm\mgd0.24$, imply that the temperature of 
dust contributing to 
the near-infrared emission of the center must be between 
$800\,\rm K\null<T_{\rm d}<1400\,$K. Lower temperatures leave no 
room for a significant dust contribution at $K$, and higher 
temperatures produce colours too blue in ($K-L$) and ($K-M$). We thus 
conclude that $T_{\rm d}=900\pm100\,$K. A similar result was 
obtained by Armus et al. (1994) who concluded from their $L'$ and 
$11.7\mu$m measurements to the presence of significant amounts of 
hot dust with $300\,\rm K\null<T_{\rm d}<1000\,$K; in this respect 
we also note the absence of cold dust emitting at far-infrared 
wavelengths throughout the inner 1.5 kpc (Braine et al. 1997).
Finally, with radiating hot dust contributing to the near-infrared 
emission, especially at $K$, NGC~3079 exhibits characteristics similar 
to those of the Seyfert 1 galaxies in which Kotilainen $\&$ Ward 
(1994) found significant contributions from dust radiating at
temperatures of 600--1000 K.

Limits to the contribution of radiating dust to the emission at $2\mu$m 
may be estimated from the deep CO-band absorption evident in Fig.~2 of HIGW.
Comparison with unreddened late-type stellar spectra by Arnaud et al.
(1989) and Lancon $\&$ Rocca-Volmerange (1992) suggests that up to 
25--30\% of the emission from the central $3''\times3''$ may be caused by dust.
By assuming {\it i.} the intrinsic colours of the stellar population 
to be those of the typical bulge in Fig.\,5, {\it ii.} a 
constant dust temperature $T_{\rm d}=900\,$K across the nuclear 
region, {\it iii.} identical (foreground) extinction for both radiating 
dust and stars and {\it iv.} no effect of scattering, we estimate from 
the observed $J$, $H$ and $K$-band fluxes a peak extinction in the central 
$1''$ pixel of $A_V=6\mg\pm1\mg$, with a $\sim40\%$ dust contribution in 
K. The $6''\times2''$ inner molecular disk region has somewhat lower mean 
values $A_V=5\mg\pm1\mg$ and a $\sim30\%$ dust contribution to observed K. 
Both values are similar to the nuclear extinction $\tau(H\alpha)$ $\sim$ 4 
estimated from the Balmer decrement by Veilleux et al.\ (1994), which 
provides further support for our conclusion that the extinction occurs 
mostly or entirely in the foreground. If the radiating dust suffers less 
extinction than the stars, these conclusions remain unchanged, but the 
dust contribution to the {\it emitted} (deredenned) radiation will be 
proportionally lower.

\begin{sloppypar}
The dereddened peak {\it stellar\/} surface brightness is of the order of 
$\sigma _{K}=2\times10^{-4}\,$W\,m$^{-2}\,\mu$m$^{-1}$\,sr$^{-1}$. Further 
analysis shows that the stellar light distribution has a halfwidth of 2$''$ 
perpendicular to the major axis, and 3$''$ along the major axis. Within 
the errors, the extent of the hot, radiating dust emission is the same, 
but its centroid appears displaced from the stellar centroid by $\secd 0.5$ 
to the west. The integrated hot dust emission, corrected for extinction, is 
$F_{K}=3\pm1.5\times10^{-14}$W\,m$^{-2}\,\mu$m$^{-1}$, corresponding to a 
luminosity $4\pi D^{2}\lambda F_{\lambda}=6\pm3\times10^{8}\,L_{\odot}$. 
This is $8\pm4\%$ of the far-infrared luminosity of the nucleus and $0.3\%$ 
of the mechanical luminosity of the modelled nuclear wind (HIGW). 
If the dust extinction is lower, these values decrease proportionally.
\end{sloppypar}

\begin{figure}[t]
\vspace{1.0cm}
\caption[]{Position-velocity diagrams of H$_{2}$ emission along lines parallel
to the total H$_2$ major axis (${\rm PA} = 157^{\circ}$), integrated
over 1$''$ strips perpendicular to the major axis. South is at left 
(negative position offsets), north is at right (positive offsets); 
zero position is that
of peak integrated H$_{2}$ emission. The middle diagram is along the line 
passing through both the H$_{2}$ and the 2.1\,$\mu$m continuum peaks, i.e.,
through the midplane of the disk. The upper figure is along a line
offset from the midplane towards the east by $\secd 1.5$ (projected distance 
130\,pc), and the lower figure along a line offset from the midplane to the 
west by an equal amount. H$_{2}$ contour levels are 1, 2, 3, 4, 5, 6, 
8, 10, 12.5, 15 and 17.5 in units of 
10$^{-8} $W\,m$^{-2}$\,sr$^{-1}$.}\label{fig.H2PV}
\end{figure}

\subsubsection{H$_{2}$ kinematics}\label{sec.H2kin}

\begin{sloppypar}
In Fig.\,6 
we show three position-velocity diagrams along lines parallel to the 
major axis of the total H$_2$ distribution at ${\rm PA} = 157^{\circ}$,
and offset from one another by $\secd 1.5$. This diagram indicates, again, 
the presence of two H$_{2}$ emission components: a bright central component 
with a large velocity width superposed on weaker emission which clearly 
shows rotation with approaching velocities north-northwest of the nucleus. 
The bright H$_{2}$ component can be identified with component C1 in the 
$\HI$ and OH absorption maps published by Baan $\&$ Irwin (1995), while 
the weaker emission corresponds to their component C2. Their component 
C3 has no counterpart in our images.

The bright central component has a FWHM diameter along the disk of 
$\secd 2.6$ (225\,pc) and a broad, flat-topped velocity distribution which 
agrees well with the H$_{2}$ $v=1{\to}0$ S(1) spectrum obtained by HIGW, 
who found a trapezoid shape for the line with ${\rm FWHM} = 560\kms$. 
The H$_{2}$ distribution in the lower two panels of Fig.\,6 does not
show evidence for the velocity gradient of $6.25\kms$\,pc$^{-1}$ 
($500\kms$\,arcsec$^{-1}$) assigned to component C1 by Baan $\&$ Irwin. 
If anything, the H$_{2}$ data suggest a much faster rotation of order 
$15\kms$\,pc$^{-1}$ in the {\it opposite} sense, i.e. velocity higher 
in the north, lower in the south. We suspect that Baan $\&$ Irwin may 
have been misled by the blending of the bright component C1 with the 
weaker extended component C2 (see their Fig.\,3 and see also Fig.\,2 of 
Veilleux et al.\ (1994). 
\end{sloppypar}

A qualitative estimate of the velocity width of the S(1) line as a 
function of position can be obtained by dividing the total H$_{2}$ 
emission by the central H$_{2}$ ``channel'' only. This shows 
that the H$_2$ velocity range is smallest in the midplane of the disk, 
and increases away from the disk towards the east (i.e. in the 
direction of the conical outflow), and also somewhat to the west.
The situation in NGC\,3079 appears to be similar to that in the central 
region of NGC\,4945, where Moorwood et al.\ (1996) found that the H$_2$ 
emission covers the surface of a hollow outflow cone coated on 
the inside with H$\alpha$ emission, that presumably plays a role in the 
collimation of this outflow.

\begin{sloppypar}
The more extended H$_2$ component, which is seen at upper left and lower 
right in the central panel of Fig.\,6, appears to rotate in the regular 
sense, with a velocity gradient of $1.2\kms$\,pc$^{-1}$ 
($105\kms$\,arcsec$^{-1}$). This is somewhat steeper than the gradient 
of the rigidly rotating CO component (Fig.\,5b in Sofue $\&$ Irwin 1992) 
which extends to about 10$''$ from the nucleus, but it is identical to 
the gradient determined by Baan $\&$ Irwin (1995) for component C2. 
The H$_{2}$ position-velocity diagram in the top panel of Fig.\,6 
(east of the midplane) repeats this pattern for the now less bright 
extended emission. In contrast, only very weak extended emission is 
present in the position-velocity diagram offset to the west (Fig.\,6, 
lower panel), where extinction must be considerably higher, especially 
if part of the emitting H$_{2}$ is outside the midplane of the galaxy. 
The rotation of the extended component implies a dynamical mass of 
$4\times10^{9}\,M_{\odot}$ within $230\,$pc from the nucleus, or 
$80\,M_{\odot}$\,pc$^{-3}$.
\end{sloppypar}

\subsubsection{Structure of the central region}

\begin{table*}[t]
\caption[]{Sizes of emitting components in the central region of
NGC\,3079}
\label{tab.sizes}
\begin{flushleft}
\begin{tabular}{llccl}
\hline
Component 		& Tracer & \multicolumn{2}{c}{FWHM diameter$^{a}$} & Reference \\
          		&        & \multicolumn{2}{c}{major axis $\times$ minor axis} & \\
          		&        & ($''\times''$) & ([${\rm pc}\times{\rm pc}$)         & \\
\hline
Inner region 		& bright H$_{2}$ emission 	& $2.7\times0.9$ & $235\times80$  	& this paper \\
(Cavity)       		& hot dust	    	      	& $2.9\times2.0$ & $250\times175$ 	& this paper \\
            		& CO ``hole'' 	      		&  2.5	       	 & 215         	  	& Sofue $\&$ Irwin (1992) \\
            		& stars ($K$-band)	      	& $2.8\times2.2$ & $245\times190$ 	& this paper \\
\hline
Inner molecular disk 	& diffuse H$_{2}$ emission 	& $6.0\times2.5$ & $515\times220$ 	& this paper \\
             		& CO core		    	& $7.0\times4.5$ & $610\times390$ 	& Sofue $\&$ Irwin (1992) \\
            		& strong extinction     	& $7.0\times5.0$ & $610\times430$ 	& this paper \\
             		& warm dust peak (0.8\,mm) 	& $<8$    	 & $<700$	  	& HIGW \\
\hline
Outer molecular		& extended CO emission		& $25\times10$   & $2200\times870$ 	& Sofue $\&$ Irwin (1992) \\
              		& extended moderate extinction$^{b}$ & $14\times10$ & $1200\times870$ 	& this paper \\
              		& extended warm dust$^{c}$ 	& $21\times <8$  & $1850\times <700$ 	& HIGW \\
			& idem				& $17\times 10$  & $1480\times 870$	& Braine et al. (1997) \\
\hline
\end{tabular}
\end{flushleft}
$^{a}$ corrected for extinction and finite resolution \\
$^{b}$ approximate values \\
$^{c}$ after subtraction of ``ridge'' component \\
\end{table*}

We are now in a position to combine the structural information from
various observations. In Table~2 we have collected the available
size information. In the data discussed so far, three significant  
scale sizes can be identified.

\begin{enumerate}

\item
The inner region appears to be a cavity filled with bulge stars and edges 
traced by the bright H$_2$ and hot dust emission. It corresponds to the 
$\sim 120\,$pc radius hole in the CO emission noted by Sofue \& Irwin (1992). 
As HIGW noted, such a cavity is required by the model of Duric $\&$ Seaquist 
(1988), in which it represents the central volume swept clear of molecular 
material and dust by the strong outflow from the galaxy nucleus. At the 
interface, rapid rotation prevails. Both the near-infrared images and the 
H$_{2}$ channel maps suggest that the inner outflow of NGC\,3079 contains 
excited molecular hydrogen and hot dust, presumably swept away from the 
inner molecular disk by the impacting winds.

\begin{sloppypar}
We propose that the relatively intense H$_{2}$ and hot dust emission both 
arise as the result of the impact of the nuclear outflow on dense and 
dusty molecular material at the interface between the central cavity 
and the molecular disk. The observed integrated intensity of the bright 
$v=1{\to}0$ S(1) H$_{2}$ emission component alone is 
$4.5\times10^{-17}$\,W\,m$^{-2}$; correction for extinction raises this 
value to $\sim 7\times10^{-17}$\,W\,m$^{-2}$, i.e. to a luminosity 
$L_{\rm S(1)}$ = $7\times10^{5}\,L_{\odot}$, which in turn suggests a 
luminosity in all H$_{2}$ lines of about $10^{7}\,L_{\odot}$. The more 
extended diffuse H$_{2}$ has a dereddened luminosity about half that of 
the bright emission region. Thus, the molecular hydrogen luminosities we 
derive here are about a quarter of those found by HIGW, partly because of 
our lower observed value and partly because of a lower derived extinction. 
This relaxes the already low efficiency requirements discussed by HIGW even 
further, so that there can be no doubt that the impacting winds can indeed 
easily explain the observed molecular hydrogen emission. As the estimated 
total H$_{2}$ luminosity of about $1.5\times10^{7}\,L_{\odot}$ is forty
times lower than the hot dust luminosity of $6\times10^{8}\,L_{\odot}$,
the latter obviously poses a more critical efficiency constraint than the 
H$_{2}$ luminosity. Although the dust efficiency requirement appears to be 
compatible with models of the type proposed by Draine (1981) it is, however, 
difficult to quantify this in the absence of further data. 

\item
The inner molecular disk extends to a radius of about 300\,pc, where its 
thickness has increased from $70-150\,$pc to about 400\,pc, suggesting an
opening angle of about 110$^{\circ}$. Excited H$_{2}$ and hot dust are 
found throughout the disk but at intensities much reduced from those at 
the interface. 
\end{sloppypar}

\item
More extended emission from warm dust and molecular gas traces the
cooler outer parts of the molecular disk out to radii of about 1\,kpc, 
after which the emission merges with the low-level emission from the 
main body of the galaxy (see Braine et al. 1997). This cooler material 
extends to distances of about 400\,pc from the plane of the galaxy.
\end{enumerate}

\subsection{Molecular gas in NGC\,3079}

\subsubsection{Relation of CO emission to H$_{2}$ column density}

\begin{sloppypar}
We may connect the total hydrogen column density $N_{\rm H}$ to reddening
and CO intensity by the following relations:

\begin{equation}
{N_{\rm H} = N{\rm(HI)} + 2N{\rm(H_{2})} 
= 5.8\times10^{21}\,f_{\rm g}\,E(B-V)}
\end{equation}

and:

\begin{equation}
{N({\rm H}_2) = 2\times10^{20}\,f_{\rm x}\,I({\rm CO})}
\end{equation}

\noindent
where $f_{\rm g}$ and $f_{\rm x}$ are the factors by which respectively
the gas-to-dust and the CO intensity to H$_{2}$ column density ratios in 
the center of NGC\,3079 differs from the canonical values; $E(B-V)$ is in
mag, $I(CO)$ is in K\,km\,s$^{-1}$ and $N$ is in cm$^{-2}$. 
Our choice of both $f_{\rm x}$ and 
$f_{\rm g}$ is such that they will be less than unity in environments with 
metallicities higher than those in the solar neighbourhood.

Baan $\&$ Irwin (1995) derive for their extended component C1 an $\HI$ 
absorption column density $N(\HI) = 3.3 \times 10^{19}\,T_{\rm s}$, where 
$T_{\rm s}$ is the unknown $\HI$ spin temperature. For the same extended 
region, we find a mean extinction $A_{V} = 5\mg$ corresponding to 
$E(B-V) = \mgd1.6$. Considering that extinction and HI absorption sample 
only half of the line of sight sampled by CO emission, we obtain:

\begin{equation}
{3.3\times10^{19}\,T_{\rm s} + 
2\times10^{20}\,f_{\rm x}\,I({\rm CO})\,=\,9.3\times10^{21}\,f_{\rm g}}
\end{equation}

\end{sloppypar}

\noindent
>From Young et al.\ (1988) we find that the central 8$''$ (695\,pc) yields 
a $J$=1--0 CO emission signal $I({\rm CO})\approx 880\,$K\,km\,s$^{-1}$,
so that:

\begin{equation}
{f_{\rm g} = 0.36\,{T_{\rm s}\over 100\,{\rm K}} + 19\,f_{\rm x}}
\end{equation}

First, we obtain from this equation an upper limit to the spin temperature 
$\HI$ associated with the extended component C1 by assuming zero H$_{2}$ 
column density towards the center: $T_{\rm s}<275\,f_{\rm g}$\,K . 
If the gas-to-dust ratio in the centre of NGC\,3079 is less than that in 
the solar neighbourhood ($f_{\rm g} < 1$), the limit on $T_{\rm s}$ becomes 
more stringent. Second, since $T_{\rm s}>0$, we find 
$f_{\rm x}<0.05\,f_{\rm g}$, i.e., even for a `normal' gas-to-dust ratio
($f_{\rm g}$ = 1) CO luminosities in the centre of NGC\,3079 correspond 
to at most a twentieth of the H$_{2}$ column density we would obtain 
by applying the `standard' Galactic conversion factor. The conversion factor
appropriate to NGC\,3079 is thus 
$X_{\rm NGC\,3079}\,\leq1\,\times10^{19}\,f_{\rm g}\,I({\rm CO})$. 
This value is also an upper limit because the value of $I({\rm CO})$ 
used here applies to a larger area than used for the extinction.
For low gas-to-dust ratios (i.e., $f_{\rm g}<<1$) and reasonable spin 
temperatures, $f_{\rm x}$ and consequently $X_{\rm NGC\,3079}$ rapidly 
become small. Conversely, CO-to-H$_{2}$ conversion factors similar to 
that of the Galactic disk ($f_{\rm x} = 1$) are obtained only for very 
{\it large\/} gas-to-dust ratios ($f_{\rm g}>20$). Such large gas-to-dust
ratios are more characteristic for extremely metal-poor dwarf galaxies 
than for the centres of spiral galaxies. 
These results are rather constrained and point to low values of both the 
$\HI$ spin temperature and the CO-to-H$_{2}$ conversion factor $X$ for a 
large range of acceptable gas-to-dust ratios. 

A low value for the CO-to-H$_{2}$ conversion factor is consistent with 
the `discrepancies' between H$_{2}$ masses derived from CO and from
submillimetre observations, noted in Sect.~5 of HIGW. Moreover, a rather
similar result has been derived by Braine et al. (1997). On the basis of
their $1.2 mm$ observations, they arrive at a {\it conservative estimate}
$X_{\rm NGC\,3079}\,\approx\,3\times10^{19}\,f_{\rm g}\,I({\rm CO})$. 
There is some evidence that within the cavity, the {\it nucleus itself} 
is surrounded by a high-density, parsec-sized accretion disk. Baan $\&$ 
Irwin's (1995) component A exhibits a high value $N_{\rm H}/T_{\rm s} = 
2\times10^{20}$ cm$^{-2}$\,K$^{-1}$. Trotter et al. (1998) argue that 
this component is part of an inner jet emanating from a heavily absorbed 
nuclear engine, and that this nucleus is surrounded by a turbulent and
presumably dense disk, 2 pc in diameter and traced by H$_{2}$O maser 
emission. The nucleus may thus suffer a much higher extinction than the 
extended region C1. This does not change our results, because both our 
extinction and the HI absorption value used do not refer to the nucleus, 
but to the material in front of the extended cavity. In addition, the 
proposed circumnuclear disk has a very small filling factor with respect 
to the 8$''$ region sampled in CO emission.

Even if the actual extinction were to be higher than assumed by us,
we would {\it still require the conversion factor $X$ to be much lower} 
than the one applicable to the solar neighbourhood, although the 
constraints on spin temperature would be much relaxed. Finally, we test 
our result by calculating the extinction towards the extended central 
region required by Eq.~(3) if $f_{\rm x}$ and $f_{\rm g}$ are both set 
to unity (i.e., Galactic values for the CO-to-H$_2$ conversion factor 
and the gas-to-dust ratio). In this case Eq.~(3) changes to the expression 
$E(B-V)=0.0057\,T_{\rm s}+30$, so that $E(B-V)\approx30$ and $A_V\approx95$. 
This value of the extinction, although possibly appropriate to the
very nucleus, is clearly ruled out for the extended circumnuclear region
sampled by our data. This underlines the robustness of our conclusion that 
at least the CO-to-H$_2$ conversion factor in the centre of NGC\,3079 is 
substantially lower than the Galactic value.

The nuclear activity in NGC\,3079 may be responsible for the apparent,
extremely low [H$_{2}$]/[CO] abundance. Theoretical models by Neufeld
$\&$ Dalgarno (1989) predict that this may occur in regions exposed
to dissociative shocks. Behind the shock, most of the carbon will be
incorporated into CO by gas-phase reactions, but the catalytic formation 
of H$_{2}$ is severely inhibited if the essential dust grains are heated
to temperatures of the level we propose for the inner parts of NGC\,3079.
As a result, [H$_{2}$]/[CO] abundances may be depressed by one to two
orders of magnitude. These theoretical predictions at least provide a 
consistent framework for the interpretation of the phenomena observed
in the central region of NGC\,3079: the presence of fast, energetic 
nuclear winds, shocked molecular hydrogen, hot dust and an underabundance
of molecular hydrogen with respect to carbon monoxide.

\subsubsection{Gaseous content of NGC\,3079}

\begin{sloppypar}
Comparison of the interferometric and single dish CO data suggest that of 
the order of 35 to 50\% of the CO emission from NGC\,3079 finds its origin 
in the inner molecular disk within 500\,pc from the nucleus (Young et al.\ 
1988). However, in the preceding we have presented evidence for a 
relatively small amount of molecular hydrogen associated with the
bright central CO emission. As the main body of NGC\, 3079 may well be 
characterized by more ``normal'' CO-to-H$_{2}$ conversion factors (e.g.
Braine et al. 1997) we cannot conclude that most of the molecular mass 
is concentrated in the central region. 

Scaling the molecular mass estimate by Young et al.\ (1988) to our adopted 
distance and our estimated CO-to-H$_{2}$ conversion factor, we find for 
the molecular disk a mass $M({\rm H}_{2}) = 8-12\times10^{7}\,M_{\odot}$, 
or $2-3\%$ of the dynamical mass. For $T_{\rm s}=100\,$K and 
$X = 1\times10^{19}$, we find a mean ratio $M({\rm H}_{2})/M(\HI) = 2.7$,
characteristic for an ISM dominated by molecular clouds.
For the rest of the galaxy, Young et al.\ found a mass of 
$8-10\times10^{9}\,M_{\odot}$ which reduces to $2-3\times10^{9}\,M_{\odot}$ 
after scaling to our adopted distance and a ``normal'' conversion factor of 
$2\times10^{20}$. This is 25 times the central molecular mass, and about a 
third of the $\HI$ mass of NGC\,3079 (Irwin $\&$ Seaquist 1991). Including 
helium, the total mass of gas in NGC\,3079 is $1.5\times10^{10}\,M_{\odot}$, 
or about 10$\%$ of its total mass (cf. Irwin $\&$ Seaquist 1991). Such a 
fraction of the total mass is characteristic for late-type galaxies, as is 
the mean ratio $M({\rm H}_{2})/M(\HI) = 0.3$. We note that our molecular 
gas estimates again are very similar to those derived by Braine et al. (1997) 
under different assumptions.

It thus appears that the molecular hydrogen content of NGC\,3079 is 
{\it not exceptional} compared to that of other late-type galaxies.
Rather, the emissivity of CO in the central molecular disk is unusually 
high.
\end{sloppypar}

\section{Conclusions}

\begin{enumerate}

\item 
The central 20$''$ ($R = 1.75\,$kpc) of NGC\,3079 exhibits a large range
of near-infrared colours, representing a varying combination of intrinsic
stellar colours, scattered stellar light, emission by hot dust and extinction 
increasing towards the nucleus. As a consequence, proper interpretation of the
observed light in terms of nuclear structure and composition cannot be
achieved by the use of photometry in multi-arcsec apertures, but requires
imaging at the highest possible resolution.

\begin{sloppypar}
\item
The nucleus suffers from significant extinction, even at near-infrared 
wavelengths. The peak extinction at a resolution of 1$''$ is 
$A_{V}=6\mg\pm1\mg$. The mean extinction of the inner $6''\times2''$
disk is $A_{V}= 5\mg\pm1\mg$. 
\end{sloppypar}

\item 
The eastern part of the NGC\,3079 bulge has ($J-H$) colours too blue
to be explained by stars in a typical quiescent bulge, 
and provide evidence either for a 20\% contribution of directly emitted 
light from young stars in the bulge, or a 20--30\% contribution by 
scattered light from stars in the bulge or in the stellar disk. 
Scattering of power-law ($S_{\nu}\propto\nu$) emission from a nuclear 
source is less likely as it would require a rather unusual intrinsic 
spectrum.

\item 
The $JHK$ colours show the presence of two or three dark lanes 
obscuring stellar light west of the nucleus. These dust lanes 
cause significant extinction of both bulge and disk.

\item 
The colours of the central 3$''$ ($R = 260\,$pc) are extremely red, 
peaking at $(J-H) = \mgd 1.5$ and $(H-K) = \mgd 1.2$. They can 
be explained by the presence of hot dust in the central region, radiating 
at temperatures close to $1000\,$K\@. The $K$-band luminosity of this hot 
dust is at most $8\pm4\%$ of the far-infrared luminosity from the central 
region, and $\sim3\%$ of the mechanical luminosity that appears to be 
available from nuclear winds in the central region.

\item
Molecular hydrogen $v=1{\to}0$ S(1) emission originates in a compact source
centred on the nucleus and elongated along the major axis, surrounded by
a region of lower surface brightness. East of the nucleus, some H$_{2}$ 
emission appears associated with the inner outflow seen at radio and optical 
wavelengths. It may represent material swept away from the molecular disk 
out of the plane by the impacting winds. A western counterpart is lacking, 
and the sharp cutoff of H$_{2}$ emission testifies to the significant 
near-infrared extinction caused by the galaxy disk intervening 
in the line of sight. 

\item 
The distribution of hot dust emission is practically identical to that
of the bright molecular hydrogen emission and just covers the central 
cavity observed in the CO distribution. This morphology, and the kinematic 
information obtained from the H$_{2}$ images, is supporting evidence for 
the nuclear wind model proposed by Duric $\&$ Seaquist (1988) and for the 
conclusions reached by Veilleux et al.\ (1994) and Hawarden et al.\ (1995). 
The H$_{2}$ and hot dust emission appears to originate in dense material 
shocked by fast nuclear winds impacting at a radius of about 120\,pc on 
the inner edge of a central molecular disk.

\begin{sloppypar}
\item 
The high-density inner molecular disk extends out to a radius of about 
290\,pc. Cooler dust and molecular gas extend out to a radius of at least 
1\,kpc. 
\end{sloppypar}

\item 
The combination of extinction, $\HI$ absorption optical depth and CO 
emission places upper limits on both the $\HI$ spin temperature and 
the CO-to-H$_{2}$ conversion factor $X$ in the central region of NGC\,3079. 
For gas-to-dust ratios comparable to those in the solar neighbourhood, 
the $\HI$ spin temperature is well below 250\,K, while $X_{\rm NGC\,3079}$ 
is less than $0.05\,X_{\rm Gal}$. Notwithstanding the concentration 
of CO emission in the center of NGC\,3079, the central regions contain only
a small fraction of all molecular hydrogen in the galaxy. The molecular
hydrogen content of NGC\,3079 is similar to that of other late-type galaxies,
but the centrally concentrated CO appears unusually overabundant with
respect to H$_{2}$, possibly related to the nuclear activity.
\end{enumerate}

\begin{acknowledgements}
We thank Markus Blietz and Alfred Krabbe for their kind assistance with 
the FAST observations. The William Herschel Telescope was operated 
by the Royal Greenwich Observatory in the Spanish Observatorio del 
Roque de los Muchachos of the Instituto de Astrofisica de Canarias. 
The research of P.P.~van der Werf has been made possible by 
a fellowship of the Royal Netherlands Academy of Arts and Sciences.
\end{acknowledgements}

\end{document}